\documentstyle[ltwol,epsfig]{article}

\def\ra{\rightarrow}
\def\be{\begin{equation}}
\def\ee{\end{equation}}
\def\bea{\begin{eqnarray}}
\def\eea{\end{eqnarray}}
\def\GeV{{\rm GeV}}

\begin{document}

\title{\vspace*{-4\baselineskip}
\hfill {\rm UR-1557}\\
\hfill {\rm ER/40685/926}\\
\hfill {\rm November 1998}\\
\vspace*{\baselineskip}
GLUON RADIATION IN TOP MASS RECONSTRUCTION:\\ EFFECT OF HADRONIC
W DECAYS}

\author{L.H.~ORR}

\address{Department of Physics and Astronomy, University of
Rochester, Rochester NY 14627-0171, USA\\E-mail: orr@pas.rochester.edu}   

\author{T.~ANDRE}

\address{Department of Physics and Astronomy, University of
Rochester, Rochester NY 14627-0171, USA and\\ Department of Physics,
University of Chicago, Chicago IL 60637, USA\\E-mail: troy@hep.uchicago.edu}   

\author{T.~STELZER}

\address{Department of Physics, University of Illinois, Urbana, IL 
61801, USA\\E-mail: stelzer@pobox.hep.uiuc.edu}

\twocolumn[\maketitle\abstracts{ Top quark events in hadron collisions 
often contain additional hadronic jets
from gluon bremsstrahlung off the quarks and gluons in the hard
subprocesses.  Such extra jets must be taken into account in attempts to
reconstruct the momentum of the top quark from those of its decay products.
We have performed a complete calculation of gluon radiation in top
production and decay at the Fermilab Tevatron including
hadronically decaying W bosons.  
In this talk we discuss the effect of gluon radiation on the
reconstructed mass of the top quark, using various top mass reconstruction
scenarios.  Implications for  the LHC are briefly 
discussed.\footnote{Presented by LHO at the 
XXIX International Conference on High Energy Physics, Vancouver, B.C.,
July 23-29, 1998.} 
}]

\section{Introduction}

Measuring the top quark mass at hadron colliders requires reconstructing
its momentum from its decay products.  Radiated gluons in top events
can complicate the reconstruction process, because for example the 
jets from gluons can be indistinguishable from the jets in top decays. 
It is important to account  correctly for these  gluons because
future top mass measurements will be dominated by systematic effects 
due to gluon radiation.

Given a top event with an extra jet from a radiated gluon, what
should we do with the extra jet?  In particular, 
 should the extra jet be combined with the $W$ and 
$b$ quark to reconstruct $m_t$?  The answer depends on where the gluon
originated.  If it was radiated from an initial state quark,
then it is a correction to the production process that is not part
of the top decay, and it should be ignored.  If the gluon was radiated from one
of the $b$ quarks from the $t$ or $\bar t$ decay, then it is itself part
of the decay and should be included in the reconstruction.  Suppose the
gluon was radiated by the top quark itself -- is it  associated with top
production or decay?  In fact it can be either, depending on when the 
top quark went on shell.  

The point is that in a given event we cannot usually distinguish between the 
possibilities (even apart from the fact that they interfere), so
we must consider top production and decay together in our treatment of
gluon radiation.  This has been done for top production and decay
at the Tevatron~\cite{OSS} and LHC~\cite{OSSLHC},  without radiation
from hadronic decays of the $W$ bosons.  But in the  detection modes in
which at least one of the top quarks can be fully reconstructed
from its  decay products  --- the lepton $+$
jets and all-jets modes ---  one or both of the $W$ bosons decays to 
quarks, which can themselves radiate.  Radiation from hadronically decaying
$W$'s in top events was treated in the soft gluon approximation in~\cite{MOS}.
The soft approximation serves as a useful guide to the distribution
of gluons and the relative importance of the various contributions, but 
it does not incorporate exact kinematics and cannot be used to study
mass reconstruction.

In this talk we present the results for an exact calculation of gluon 
radiation in top production and decay at the Tevatron 
with hadronic $W$ decays fully taken into account.

\section{Gluon Distributions}

We have calculated the cross section for 
$p\bar p\ra b\bar b q \bar q\prime l \nu j$ and 
$p\bar p\ra b\bar b q \bar q\prime  q \bar q\prime j$ where $j$ is an
extra radiated jet.  This tree-level calculation is exact at 
${\cal{O}}(\alpha_s^3)$ and contains all spin correlations, top width effects,
and interferences.  The center of mass energy is 1.8 TeV, and top and 
bottom masses are 175 and 5 GeV, respectively.
Helicity amplitudes are computed with the assistance
of the MADGRAPH package~\cite{MADGRAPH}.  The results shown below
are for  the  $q\bar q$ initial state that dominates in top production
at the Tevatron; we have done the calculation for  the $gg$ and $qg$ 
initial states~\cite{AOS} but do not show those results here.

We apply the following kinematic cuts to all final-state jets (which
in this parton-level calculation are quarks and gluons) and to the
charged lepton:
\bea
E_{Tj}, E_{Tl} \> & \geq & \>  15 \ {\rm GeV} \; ,\nonumber \\
|\eta_j| , |\eta_l| \> & \leq & \> 2.5 \; ,\nonumber \\
\Delta R_{jj}, \Delta R_{jl} \> & \geq & \> 0.4 \; .
\label{cuts}
\eea
These are meant to mimic experimental cuts, so that the partons are likely to
appear in the detectors with enough angular separation to be distinguishable
as separate particles.  They also protect the theoretical cross section from
the soft and collinear singularities that appear at tree level.

In the distributions we present below we will decompose the cross section
into contributions from radiation associated with various parts of the 
process.  These contributions are:\\
{\bf i.} {\it Production-stage radiation,} which comes
from the initial quarks or internal gluon line, or from the top (or antitop)
quark {\it before} it goes on shell.  \\
{\bf ii.} {\it Decay-stage radiation} that is part of the $t$ or $\bar t$
decay; this is further subdivided into

{\it a. Decay-tb radiation} from either of the $b$'s or from either 
of the $t$'s {\it after} they go on shell.

{\it b. Decay-W radiation} from the decay products of hadron\-ically decaying 
$W$ bosons. 

We  make these distinctions in the parton-level calculation based 
on kinematics to see how the various contributions behave; 
this cannot of course be done for a given event in the experiments.
In principle the production-stage and decay-stage contributions
can interfere with each other (with the exception of the 
decay-$W$ radiation, which cannot interfere with the other processes
because the $W$ is a color singlet).  And although we do include all 
interferences in our calculation, 
in practice the production-decay interference is very small for  gluon
energy thresholds large compared to the top width $\Gamma_t=1.5\ \GeV$,
as in the present case.

\begin{figure}
\psfig{figure=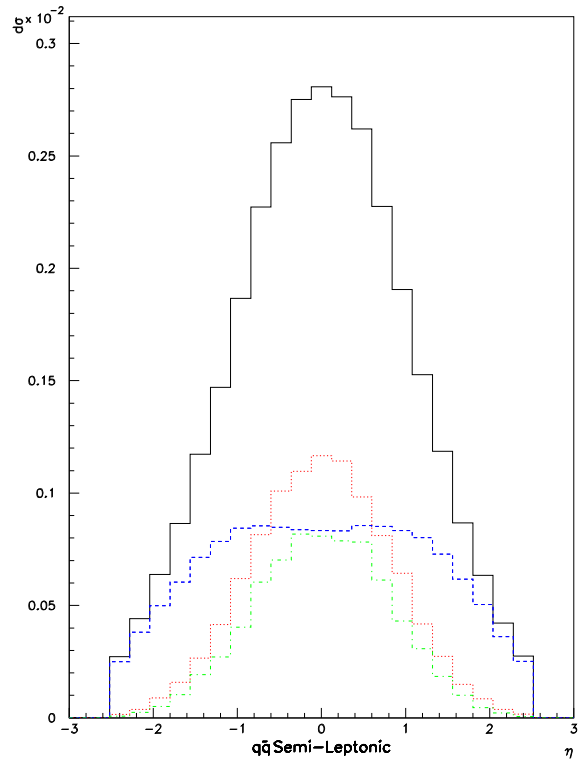,width=8cm}
\caption{The extra jet pseudorapidity  ($\eta_j$) distribution
(solid histogram) and its
decomposition in terms of production (dashed
histogram), decay-$tb$ (dotted histogram) and decay-$W$ (dot-dashed histogram) 
emission contributions, for the lepton $+$ jets mode.
}
\label{fig:etalj}
\end{figure}

Figure~\ref{fig:etalj} shows the distribution in pseudorapidity of the 
extra jet at the Tevatron for the lepton $+$ jets case, {\it i.e.\/}
for a single hadronically decaying $W$.  
The production-stage radiation, shown as 
a dashed histogram, has the broadest distribution, populating
most of the accessible rapidity range.  The two decay-stage contributions
are more centrally peaked.  The decay-$tb$ contribution (dotted 
histogram) is slightly larger,
as it accounts for radiation from both the $t$ and $\bar t$, but the 
decay-$W$ contribution (dot-dashed histogram) from a single 
$W$ is similar in size and shape to the decay-$tb$, as was found in the 
soft approximation~\cite{MOS}.  The central region of the detector 
is populated by all three contributions, which means that distinguishing
them will be challenging at best.

Results for the all-hadronic mode, where both $W$ bosons decay to quarks, are 
similar, the main difference being that the decay-$W$ contribution 
approximately doubles in size.

\begin{figure}
\psfig{figure=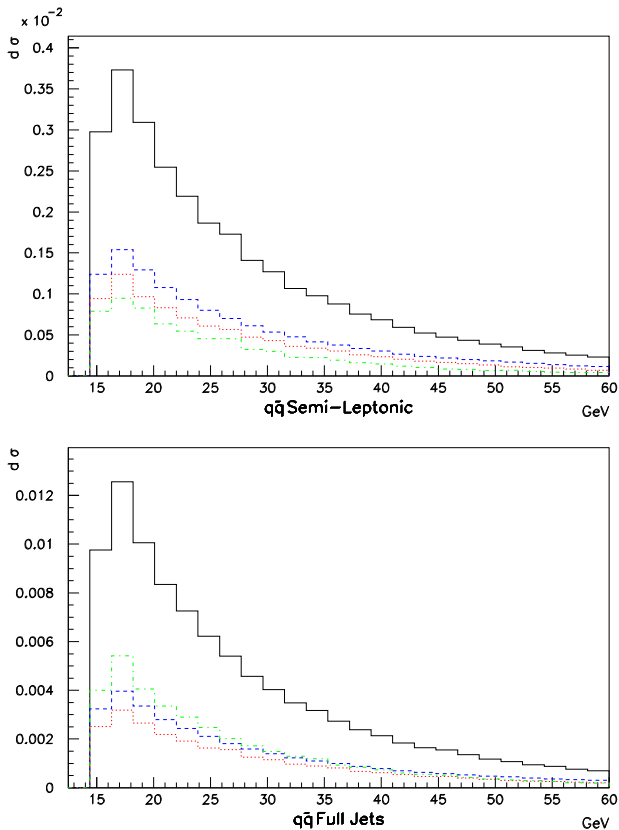,height=4.5in}
\vskip -.25cm
\caption{The extra jet transverse energy distribution
(solid histogram) and its
decomposition in terms of production (dashed
histogram), decay-$tb$ (dotted histogram) and decay-$W$ (dot-dashed histogram) 
emission contributions, for the lepton $+$ jets mode (top plot) and 
all-hadronic (bottom plot)
mode.
}
\label{fig:et}
\end{figure}

The transverse energy distributions of the radiated jet are shown in 
Figure~\ref{fig:et} for the lepton $+$ jets (top) and all-hadronic
modes.  The spectra look quite similar for the various contributions,
and extend to large values of transverse energy.

\section{Top Mass Reconstruction}

Given the difficulty of distinguishing extra jets radiated in the production
stage from those from decay, we can ask what effect the extra jet has on
mass reconstruction.  First, however, we have to ask which jet {\it is}
the extra one.  It seems reasonable to assume that the gluon jet is the 
one with the lowest $E_T$, given the infrared singularity that characterizes
emitted radiation.  While it is true that of the final state jets, the 
gluon has the softest $E_T$ spectrum, the gluon has the lowest $E_T$
only just over half the time (just under half for the all-jets mode).
Still, there is no method that is obviously better for identifying the 
extra jet.

We reconstruct the top mass from the final state partons  without assuming 
we know which jet is which, but we omit the jet with the lowest $E_T$.
For the lepton $+$ jets mode we do the following.\\
(1.) Drop the lowest $E_T$ jet.  This leaves four jets.\\
(2.) Find the jet pair with invariant mass closest to $m_W$.\\
(3.) Solve for the neutrino four-momentum using the charged lepton momentum
and $W$ mass constraint.\\
(4.)  Combine each of the $W$'s with each of the remaining jets to give 
$m_t$ and $m_{\bar t}$.  Choose the combination that minimizes the $t$-$\bar t$
mass difference.

\begin{figure*}
\psfig{figure=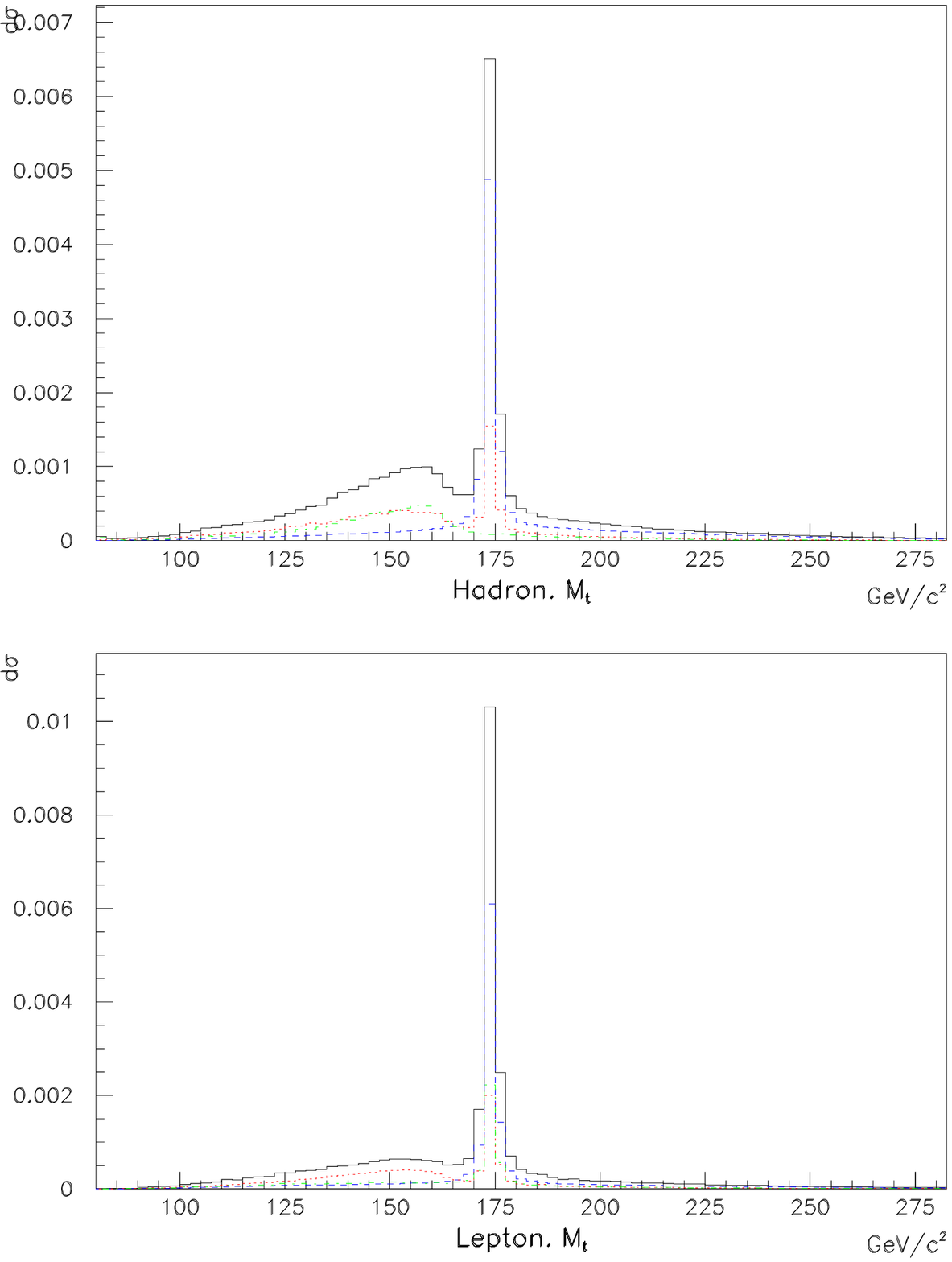,width=16cm}
\vskip -.25cm
\caption{The reconstructed top mass distributions (solid histograms) 
in the  lepton $+$ jets mode
for the hadronically decaying $W$ (top plot)
and the leptonically decaying $W$, and their 
decomposition in terms of production,
 decay-$tb$  and decay-$W$   
emission contributions.
}
\label{fig:mtlj}
\end{figure*}

The $m_t$ distributions obtained from this procedure are shown in 
Figure~\ref{fig:mtlj}.  The top plot shows the distribution
for the top with the hadronically decaying $W$, and the bottom corresponds
to leptonic $W$ decay.  In both cases we see a peak at the correct central 
value, where the procedure resulted in the correct mass. 
We also see smooth, reasonably flat high and low tails from wrong 
combinations, in addition to bumps in the low tails corresponding to 
the omission of jets that were part of the decays.  Note that these
bumps appear in the contributions from decay-stage radiation, when 
{\it all} of the jets should be included in the mass reconstruction.
Finally, the leptonically decaying $W$ gives a sharper top mass distribution
because with no radiation from the $W$ decay, there are fewer wrong
combinations.

\begin{figure}
\psfig{figure=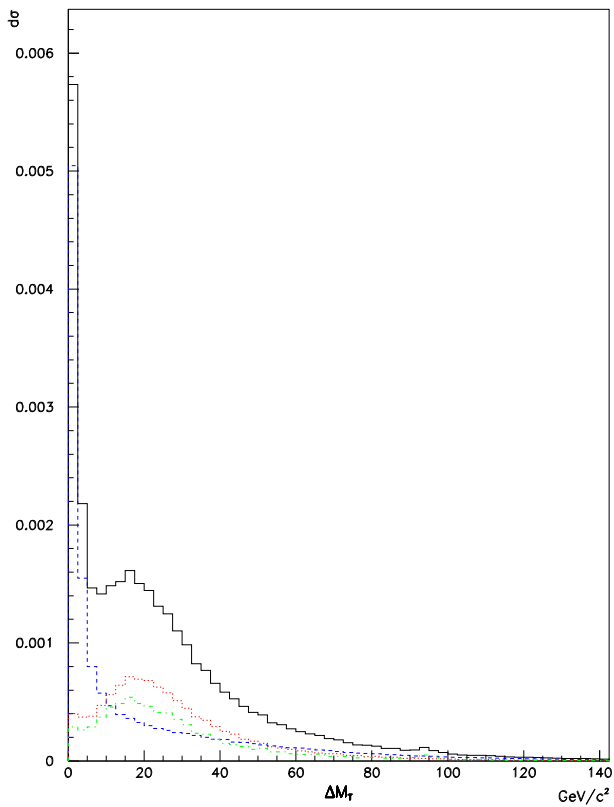,height=4.5in}
\vskip -.25cm
\caption{The distribution in magnitude of the $t$-$\bar t$ mass
difference for the events in the previous figure.
}
\label{fig:deltam}
\end{figure}

The results shown in Figure~\ref{fig:mtlj} are meant to be illustrative
and should not be taken as a direct representation of distributions
measured in experiment.  In particular, this calculation is at the 
parton level; we have not included backgrounds; and these distributions
only include events with a radiated gluon.  The effects of hadronization,
energy resolution and detector effects, and background will certainly
make things worse.  However there are certainly ways to improve on this
simplistic analysis as well.  For example, although
we minimized mass differences we did not cut on them explicitly.
Figure~\ref{fig:deltam} shows the magnitude of the $t$-$\bar t$ mass
difference on an event-by-event basis for the case shown above.  It
suggests that an absolute cut on the mass difference could reduce the 
tails in the mass distributions.  
Interestingly, $b$-tagging, {\it i.e.\/} assuming we can identify $b$ jets, 
does not improve the mass distributions much.  This is because
we do not include backgrounds.  

The results for the all-jets mode are similar, except that with both $W$'s
decaying to quarks there are two more jets in each event, leading
to more possible wrong combinations and a corresponding increase in the 
tails.

\section{Conclusions}

In top quark physics, as statistics improve, systematic effects associated 
with gluon radiation will dominate measurements of the top mass.  
We have added hadronic $W$ decays to analyses of gluon radiation in
top production and decay and presented some initial results here.
We find that the contribution from radiation from a single hadronically 
decaying $W$ is nearly as large as and comparable in shape to the remaining
decay-stage radiation from both the $t$ and $\bar t$.  The presence of 
radiation from both the top production and decay stages complicates the
reconstruction of the top momentum from its decay products and 
hence complicates the measurement of the top mass.  Further analysis
is in progress for the Tevatron and LHC~\cite{AOS}.

\section*{Acknowledgements}
Work supported in part by the U.S. Department of Energy,
under grant DE-FG02-91ER40685 and by the U.S. National Science Foundation, 
under grant PHY-9600155.

\end{document}